\documentclass[a4paper,twocolumn]{article}

\usepackage{epsfig}
\usepackage{amssymb}

\begin{document}

\title{Switching between memories in neural automata with synaptic noise}
\author{J. M. Cortes, P. L. Garrido, J. Marro and J. J. Torres  \\
Departamento de Electromagnetismo y F\'{\i}sica de la Materia, and \\
Institute \emph{Carlos I} for Theoretical and Computational Physics, \\
University of Granada, 18071--Granada, Spain}

\maketitle

{Neurocomputing 58-60: 67-71, 2004}

{Corresponding author: Jesus M. Cortes}

{mailto:jcortes@ugr.es}

\begin{abstract}

We present a stochastic neural automata in which activity fluctuations and
synaptic intensities evolve at different temperature, the latter moving
through a set of stored patterns. The network thus exhibits various
retrieval phases, including one which depicts continuous switching between
attractors. The switching may be either random or more complex, depending
on the system parameters values.
\end{abstract}

\section{Introduction and Model}

Understanding how the processing of information in neural media is
influenced by the biophysical processes that take place at the synaptic
level is an open question. In particular, the effect of synaptic dynamics
and noise on complex functions such as associative memory is not yet well
understood. In relation to this, it has been reported that short term
synaptic plasticity has a main role in the ability of some systems to
exhibit switching between stored memories.\cite{jtorresNC} The same behavior
ensues assuming dynamics of the neuron threshold to fire.\cite{dhornPRA} The
origin of the switching mechanism seems in both cases at a sort of fatigue of the
postsynaptic neuron under repeated presynaptic simulation. This destabilizes
the current attractor which may result in a transition to a new attractor.
It would be interesting to put this on a more general perspective concerning
the role of noise in associative memory tasks. With this aim, we present in
this paper a \emph{stochastic neural automata} that involves two independent
competing dynamics, one for neurons and the other for synapses.

Consider $N$ (binary) neuron variables, $s_{i}=\pm 1$, any two of
them linked by synapses of intensity $w_{ij}$; $i,j=1,\ldots ,N$. The
interest is on the configurations $\mathbf{S}\equiv \{s_{i}\}$ and $\mathbf{W
}\equiv \{w_{ij}\}$. In order to have a well--defined reference, we assume
that interactions are determined by the Hopfield \textit{energy} function.
 Furthermore, consistent with the observation that memory is a global
dynamic phenomenon, we take the model dynamics determined at each time step
by a single pattern, say $\mu $. Consequently, $H(\mathbf{S},\mathbf{W};t)=-
\frac{1}{2}\sum_{i}\sum_{j\neq i}w_{ij}^{\mu }s_{i}s_{j}$ with $\mu =\mu(t)$
and assuming the Hebbian learning rule, for example,
$w_{ij}^{\mu }=\frac{k}{N}\xi _{i}^{\mu }\xi_{j}^{\mu }$,
 where, $\xi _{i}^{\mu }=\pm 1$  are the variables that characterize the $\mu$ pattern, one out of the $P$
  \emph{memorized} ones, and $k$ is a proportionality constant.
Therefore, each configuration $\mathbf{W}$ is unambiguously associated to
a single $\mu$, and we write $\mathbf{W}\equiv \mu $ in the following.

The above may be formulated by stating that the probability of any
configuration $(\mathbf{S},\mu )$ evolves in discrete time according to
\begin{equation}
P_{t+1}({\mathbf S},\mu )=\sum_{{\mathbf S^{\prime }}}\sum_{\mu^{\prime }}
T[ (\mathbf{S},\mu) | (\mathbf{S^{\prime}},\mu^{\prime}) ] P_{t}(\mathbf{S^{\prime }},\mu ^{\prime }),
\label{discrete_master_equation}
\end{equation}
where $T[ ( \mathbf{S},\mu ) | ( \mathbf{S^{\prime }},\mu^{\prime }) ]$ represents the probability of jumping from
$(\mathbf{S^{\prime }},\mu ^{\prime })$ to $(\mathbf{S},\mu )$. We
explicitly consider here the case in which
\begin{equation}
T[ (\mathbf{S},\mu) | ( \mathbf{S^{\prime }},\mu^{\prime}) ] =
T_{0}^{\mu ^{\prime }}[ \mathbf{S} | \mathbf{S^{\prime }} ] \times T_{1}^{\mathbf{S}}[ \mu  | \mu ^{\prime} ]
\label{probability_of_jumping}
\end{equation}
with $T_{0}^{\mu^{\prime }}[ \mathbf{S} | \mathbf{S^{\prime}} ]$ corresponding
to \textit{Little dynamics}, i.e., parallel updating, so that
$T_{0}^{\mu^{\prime }}[ \mathbf{S} | \mathbf{S^{\prime }} ]=
\prod_{i=1}^{N}t_{0}^{\mu^{\prime }}[s^{\prime },i]$.
Furthermore, $t_{0}^{\mu^{\prime }}[s^{\prime },i] \equiv
\Psi [ \beta _{0}\Delta H^{\mu^{\prime }}( s_{i}^{\prime } \rightarrow s_{i}=
\pm s_{i}^{\prime}) ]$, where $\Psi(X)$ is an arbitrary function,
except that  it is taken to satisfy \textit{detailed balance}
(see Ref.\cite{jmarroBOOK} for a discussion), $\beta _{0}$
is an (inverse) temperature parameter, and $\Delta H$ denotes the \textit{energy}
change brought about by the indicated transition. For changes in
the synapses, we take $T_{1}^{\mathbf{S}}[ \mu | \mu ^{\prime } ]=\Psi[ \beta _{1}
\Delta H^{\mathbf{S}}(\mu ^{\prime }\rightarrow \mu) ]$. We
also take  $\sum_{\mathbf{S}}\sum_{\mu }T[ (\mathbf{S},\mu
) |( \mathbf{S^{\prime }},\mu ^{\prime })] =1$ for
any $(\mathbf{S^{\prime }},\mu ^{\prime })$. After some algebra, one has  that
$\Delta H^{\mu ^{\prime }}(s_{i}^{\prime } \rightarrow s_{i}=\pm s_{i}^{\prime})=
-k\xi _{i}^{\mu ^{\prime }}( s_{i}-s_{i}^{\prime })
(m^{\prime \mu ^{\prime }}-s_{i}^{\prime }\xi _{i}^{\mu ^{\prime }}/N)$ and
$\Delta H^{\mathbf{S}}(\mu ^{\prime }\rightarrow \mu )=-\frac{1}{2}kN[
( m^{\mu }) ^{2}-( m^{\mu ^{\prime }}) ^{2}]$, where $m^{\mu }(\mathbf{S})\equiv m^{\mu }$
is the overlap between the current state $\mathbf{S}$ and pattern $\mu$. The
factor $N$ in  $\Delta H^{\mathbf{S}}$ appears because we assume \emph{
global} energy variations (i.e., all synapses in the configuration are
attempted to be changed at each step) instead of the energy variation per
site in $\Delta H^{\mu^{\prime }}$.

This  model differs essentially from apparently close proposals,
 e.g., \cite{jtorresPRL,jtorresJPA,acoolenPRB}.
First, because it assumes the same time scale for changes in both $\mathbf{S}$
and $\mu$. On the other hand, the choice here for $T[ ( \mathbf{S},\mu ) | ( \mathbf{S^{\prime }},\mu ^{\prime }) ]$
amounts to drive neurons activity and synaptic intensities by
different temperature, $\beta _{0}^{-1}\equiv T_{0}$ and $\beta
_{1}^{-1}\equiv T_{1}$, respectively. The case of our
model with a single pattern is equivalent to the equilibrium
Hopfield model with $P=1$; for more than one pattern, however,
new nonequilibrium steady states ensue. This is closely due to the fact
that $T[ ( \mathbf{S},\mu ) |( \mathbf{S^{\prime }},\mu ^{\prime })]$
 does not satisfy detailed balance.\cite{jmarroBOOK}

In principle, one may estimate from (\ref{discrete_master_equation}) how any
observable $F(\mathbf{S},\mu )$ evolves in time. The result is an equation
$\langle F\rangle _{t+1}=f_{t}({\bar{K}},F),$ where ${\bar{K}}$
is the set of control parameters and $\langle \cdots \rangle $
denotes statistical average with $P(\mathbf{S},\mu )$. \cite{jmarroBOOK} 
Alternatively, one may be directly concerned with the
time evolution for the probability of jumping in terms of the overlaps $
\mathbf{m}\equiv \{ m^{\nu };\mu =1,\ldots ,P\}$ . 
One has that $\Pi _{t+1}(\mathbf{m},\mu )=\sum_{S}\delta
\lbrack \mathbf{m}-\mathbf{m(S)}]P_{t+1}(\mathbf{S},\mu )$ satisfies
\begin{equation}
\Pi _{t+1}(\mathbf{m},\mu )=\int d\mathbf{m^{\prime }}\sum_{\mu ^{\prime }}
\bar{T}[ ( \mathbf{m},\mu ) |( \mathbf{m^{\prime }},\mu
^{\prime }) ] \,\Pi _{t}(\mathbf{m^{\prime }},\mu ^{\prime }).
\label{effective_master_equation}
\end{equation}

\begin{figure}[h!]
\psfig{file=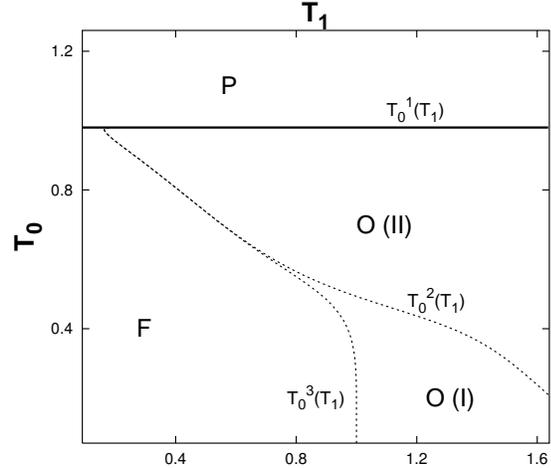,width=7.5cm}
\caption{{\small 
 Phase diagram showing three different phases. ($F$)
{\it Ferromagnetic}, for $T_{0}<T_{0}^{3}(T_{1})$, with $\mathbf{m}\neq 0$ and $
j=0$. The system has \textit{static} associative memory.
($P$)  {\it Paramagnetic}, for $T_{0}>T_{0}^{1}(T_{1})$, with $\mathbf{m}=0$ and $j=0$,
 without any kind of associative memory. ($O$) {\it Oscillatory},
for $T_{0}^{3}(T_{1})<T_{0}<T_{0}^{1}(T_{1})$, with $\mathbf{m}=0$, $j \neq 0$
 and \textit{dynamic} associative memory, e.g. there are
jumps between patterns either uncorrelated ($O(II)$) or time-correlated ($O(I)$),
 as explained in the main text. The transition between $O(I)$ and $O(II)$
  is discontinuous. Here, $N=16384$ and $P=3$ spatial-correlated
patterns with 20\% of  average overlap between any two of them. 
} }
\end{figure}

This amounts to reduce the degrees of freedom, from a number
of order $2^{N}+1$ in $(\mathbf{S},\mu )$ to $P+1$ in $(\mathbf{m},\mu ).$
Dealing with this sort of coarse--grained master equation
requires an explicit expression for $\bar{T}[ ( \mathbf{m},\mu
) |( \mathbf{m^{\prime }},\mu ^{\prime }) ] $ which
we take as \cite{acoolenDYN} $\bar{\Psi}[ \beta _{1}\Delta H^{\mathbf{m}}( \mu ^{\prime }\rightarrow \mu ) ]
 \mathcal{K}\int d \mathbf{q}\exp [ N\Phi ( \beta _{0},\mathbf{m},\mathbf{m^{\prime }},\mathbf{q},\mu ^{\prime }) ]$.
 Here, $\mathcal{K}$ is a constant, and $\mathbf{q}$ is the conjugated
momentum of $\mathbf{m}$. Hence, $\mu $ and $\mathbf{m}$  evolve
separately in time. Changes in $\mu $, given  ${\mathbf{m}}$,  are controlled by
$\bar{\Psi}[ \beta _{1}\Delta H^{\mathbf{m}
}( \mu ^{\prime }\rightarrow \mu ) ]$, while ${\mathbf{m}}$
evolves according to the term
$\int d\mathbf{q}\exp [ N\Phi ( \beta _{0} \mathbf{m},\mathbf{m^{\prime }},\mathbf{q},\mu ^{\prime }) ]$
with a fixed $\mu^{\prime }$.
A justification of this equation and a detailed study of its consequences
will be reported elsewhere.\cite{jcortesTOBE}
\section{Simulations}
Here we report on some preliminary results of a Monte Carlo study of this
model which reveals an intriguing situation. Different regimes
are shown in Figure $1$ ({\em Left}) depending on the  values of temperatures $T_{0}$ and $T_{1}$. 
To distinguish between them, we introduce the
overlap (${\bf m}$) and the total number of jumps ($j$); three regimes occur that  are
close to the ones reported in \cite{jtorresNC}.
 There is an oscillatory
phase which is illustrated in Figure $1$ ({\em Right}). The system in this case has
associative memory, like in Hopfield model. However, this is here a dynamic
process, in the sense that the system trapped in any attractor corresponding
with a pattern is able to jump to the other stored patterns. Because the
probability of jumping depends on the neurons activity, this mechanism is, in general, a complex process.
\begin{figure}[h!]
\psfig{file=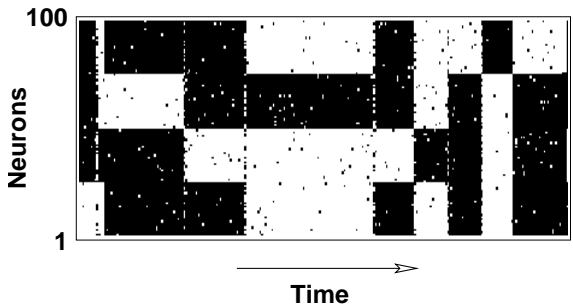,width=7.5cm}
\caption{{\small 
  Activity of neurons versus time
for $N=100$ neurons and $P=4$ patterns. Here, $T_{0}=0.9T_{0}^{c}$ and $T_{1}=1.69T_{1}^{c}$,
 where $T_{0}^{c}$ and $T_{1}^{c}$ are the
corresponding critical values of temperatures.} }
\end{figure}

One might argue that these jumps are a finite size effect;
it does not seem to be the case, however. Similar jumping phenomena,
apparently independent of the size of the system,\cite{mmunozEURLET} have already
been described in kinetic Ising--like models in which disorder is
homogeneous in space and varies with time, and mean--field solutions also
exhibit these phenomena. Some finite--size effects are evident, however;
the synaptic temperature, for instance, scales with size.  In fact, we obtain
$T_{1}^{c}/N=0.0431 \pm 0.0001$ for $N=1024,1600$ and $4096$; consequently,
  we redefine $\beta _{1}\equiv \beta _{1}N$ from now on.

A series of our computer experiments concerned $N=65536$ and $
P=6.$ In order to study in detail the oscillatory phase, it turned out
convenient to look at time correlations. Therefore, we used correlated
patterns, namely, there was an average overlap of 20\% between any two of
the stored patterns. The goal was to detect non--trivial correlations
between jumps, so that we computed the time $\tau _{\nu \gamma }$ the system
\textit{remains} in pattern $\nu $ before jumping to pattern $\gamma ;$ $
\sum_{\gamma =1}^{P}\tau _{\nu \gamma }=\tau _{\nu }$ is the total time the
system stays in pattern $\nu .$ This reveals the existence of two different
kinds of oscillatory behavior. One is such that $\tau _{\nu \gamma }\simeq
\tau ,$ independent of $\nu $ and $\gamma .$ That is, the system stays the
same time at each pattern, so that jumping behaves as a completely random process,
without any time correlation. This is denoted by \textit{O(II)} in Figure $1$ ({\em Left}).
Even more interesting is phase \textit{O(I)}. As the probability of jumping between
patterns is activity dependent, lowering $T_{0}$ leads to non--trivial time
correlations, namely, $\tau _{\nu \gamma }$ depends on both $\nu $ and $
\gamma .$ We also observe that $\tau _{\nu \gamma }$ differs from $\tau
_{\gamma \nu }$. This peculiar behavior suggests one that  some spatial temporal information may be coded
in  phase \textit{O(I)} .
\begin{figure}[h!]
\psfig{file=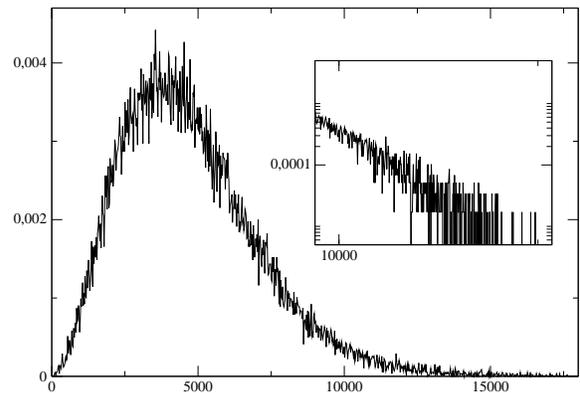,width=7cm,angle=270}
\caption{{\small  Probability distribution for the time the system stays in a pattern before jumping to
another one in the phase \textit{O(II)}. \underline{Inset:} Tail for large
events. }}
\end{figure}

In order to understand further these two different jumping mechanisms, we
simulated for $T_{0}=\{0.1,0.5,1.1\}$, in
units of $T_{0}^{c},$ at fixed $T_{1}=1.36T_{1}^{c}.$ The resulting
probability distribution of time $\tau _{\nu }$ averaged over $\nu ,$ $
P(\tau ),$ is shown in Figure $2$. The data fit $P(\tau )=A\exp (-B\tau
^{2})\tau ^{2}.$ This predicts that $\langle \tau ^{2}\rangle ^{2}=\frac{9}{
64}\pi ^{2}\langle \tau \rangle ^{4}$ which compared with our
simulations gives relative errors of $e(\%)=\{3.3,3.8,11.2\}$  for $T_{0}=\{0.1,0.5,1.1\}$,
 respectively. The error increases with $T_{0}$ because the
overlaps then tend to become too small and jumps are, consequently, not so
well-defined.

Also interesting is the average of time $\tau$ before jumping, because diverging of $\langle \tau \rangle $
indicates that the overlap is stable and no jumps occur. The {\it trial} distribution
above gives $\langle \tau \rangle =A/2B^{2}.$ Therefore, $B$, which also enters
the probability normalization as $A=4( B^{3}/\pi ) ^{1/2}$,
indicates whether there are jumps $(B\neq 0)$ or not $(B=0)$. $B$
measures the jumping frequency.

 It is also worth studying the tails of the distributions
for large events. This is  illustrated in Figure $2$ ({\em Inset}) . One may argue that,
as in Ref.\cite{phurtadoTOBE} for somewhat related
phenomena, this tail is due to the superposition of many exponentials, each
corresponding to a well--defined type of jumping event. We are presently
studying this possibility in detail.

We acknowledge P.I. Hurtado for very useful comments and financial support
from MCyT-FEDER,  project BFM2001-2841 and the J.J.T.'s \textit{Ram\'{o}n y Cajal} contract.

\end{document}